\documentstyle[psfig]{caosp}
\begin{document}
\def\teff{$T_{\rm eff}$}
\def\lgg{$\log\,{g}$}
\def\vt{$\xi_{\rm t}$}
\def\vsini{$v\cdot \sin i$}
\def\kms{\,km\,s$^{-1}$}
\def\gequ{$\gamma$\,Equ}
\def\alcir{$\alpha$\,Cir}
\def\i{\,{\sc i}} \def\ii{\,{\sc ii}} \def\iii{\,{\sc iii}}
\def\kG{\,kG}
\def\vsini{$v\cdot \sin i$} \def\hfs{hyperfine-structure\ }
%%%%%%%%%%%%%%%%%%%%%%%%%%%%%%%%%%%%%%%%%%%%%%%%%%%%%%%%%%%%%%%%%%%%%%%%%%%%%%
\pubyear{1998}
\volume{27}
\firstpage{359}
%%%%%%%%%%%%%%%%%%%%%%%%%%%%%%%%%%%%%%%%%%%%%%%%%%%%%%%%%%%%%%%%%%%%%%%%%%%%%%
\htitle{Eu\,{\sc iii} identification and Eu abundance in cool CP stars}
\hauthor{T. Ryabchikova {\it et al.}}
\title{Eu\,{\sc iii} identification and Eu abundance in cool CP stars}
\author{T. Ryabchikova \inst{1}  \and N. Piskunov \inst{2}
\and I. Savanov \inst{3} \and F. Kupka \inst{4}}
\institute{Institute of Astronomy, RAS, Moscow, Russia
\and Uppsala Astronomical Observatory, Uppsala, Sweden
\and Crimean Astrophysical Observatory, Nauchny, Crimea, Ukraine
\and Institute for Astronomy, University of Vienna, Vienna, Austria}
%
%\today
\maketitle

\begin{abstract}
We report the first identification of the Eu\,{\sc iii} $\lambda$ 6666.317 line
in optical spectra of CP stars. This line is clearly present in the spectra of
HR 4816, 73 Dra, HR 7575, and $\beta$ CrB, while it is
marginally present or absent in spectra of the roAp stars
$\alpha$ Cir, $\gamma$ Equ, BI Mic, 33 Lib, and HD 24712.
\end{abstract}

\section{Introduction}
Magnetic Chemically Peculiar stars (CP2 stars) are known to have large
overabundances of rare-earth elements (REE) in their atmospheres.
Among all REE, europium shows the most prominent overabundances of up to 
+5.0 dex, in many CP2 stars violating the odd-even pattern observed in 
the solar atmosphere. In the atmospheres of many CP2 stars the dominant 
europium ion is Eu\,{\sc iii}. The strongest lines of Eu\,{\sc iii} are located
in the UV region. A few relatively intense lines are observed in the optical 
spectral region (Sugar \& Spector 1974).
This fact justified a careful study of a few CP2 stars in the
spectral region 6620-6680 \AA, where unblended lines of both Eu\,{\sc ii}
$\lambda$ 6645.05 and the strongest optical Eu\,{\sc iii} $\lambda$ 6666.35
are located.

\section{Observations and line analysis}
\label{Obs}

A list of the programme stars is given in Table 1.
CCD spectra of five stars:  HR 4816, 73 Dra, HR 7575,
$\beta$ CrB, and GZ Lib (33 Lib) were obtained at the Crimean Astrophysical 
Observatory. For all spectra the S/N ratio is at
least 200 and the resolving power is 35000. For the remaining stars we used
spectra from our previous abundance study of roAp stars. The spectrum of the
star HD 24712 (DQ Eri) was
obtained at the Nordic observatory and kindly provided to us by V. Malanushenko.
In Table 1 a horizontal line separates a group of roAp stars (lower part)
from the non-oscillating CP2 stars (upper part).

%\newpage
\begin{table}[t]
\caption{The list of the programme stars. Effective temperatures, surface
gravities, surface magnetic fields, and rotational velocities are given.}
\label{table1}
\begin{flushleft}
\begin{small}
\begin{tabular}{lrrccccc}
\noalign{\smallskip}
\hline
 Star name    &  \teff  &  $B_{\rm s}$ & \vsini  & & log(N/H) & & log(gf)  \\
              &     &  (kG) & (km\,s$^{-1}$)&Cr & Fe  & Eu & Eu\,{\sc iii} \\
\hline
HR 4816       &   9000  & 3.6 & ~9~~   &-3.56 & -3.51 & -8.36& 1.15 \\
73 Dra        &   8900  & 2.0 & ~9~~   &-3.56 & -3.56 & -8.53& 1.25 \\
HR 7575       &   8500  & 3.6 & ~2~~   &-4.21 & -3.98 & -7.66& 1.00 \\
$\beta$ CrB   &   8000  & 5.7 &~3.5    &-4.46 & -3.96 & -8.36& 1.10 \\
\hline
\alcir        &   7900  & 2.0 &12.5    &-5.31 & -4.46 & -9.66& 0.95 \\
\gequ         &   7700  & 4.0 & ~0~~   &-5.31 & -4.31 & -9.86& 1.07 \\
BI Mic        &   7450  & 0   &12.5    &-5.51 & -4.46 & -9.96& 1.25 \\
GZ Lib        &   7350  & 4.5 &$\leq$8~&-4.86 & -4.26 & -8.81& 1.07 \\
DO Eri        &   7250  & 3.0 & ~5.6   &-5.76 & -4.96 & -9.36& ...  \\
\hline
\end{tabular}
\end{small}
\end{flushleft}
\end{table}

The Vienna Atomic Line Database (VALD, Piskunov et al. 1995) was extensively
used for line identifications, based on preliminary abundances extracted
for the program stars from the literature. VALD does not contain any information
on the second ions of the REE. A list of the Eu\,{\sc iii} lines classified by 
Sugar \& Spector (1974) was used by us. The strongest optical line Eu\,{\sc iii} 
$\lambda$ 6666.347 ($^6\rm I^0_{17/2} - ^6\rm H_{15/2}$) is very strong 
in non-roAp stars, and there are no other candidates for the 
identification of the observed feature. This line is noticeable in roAp 
stars, too, but it is very weak, and partially blended from both sides 
with unidentified lines. Figure 1 shows spectra of the programme stars 
where the positions of the Eu\,{\sc ii} ($\lambda$ 6645.05) and Eu\,{\sc iii} 
lines have been marked.

\begin{figure}
\psfig{figure=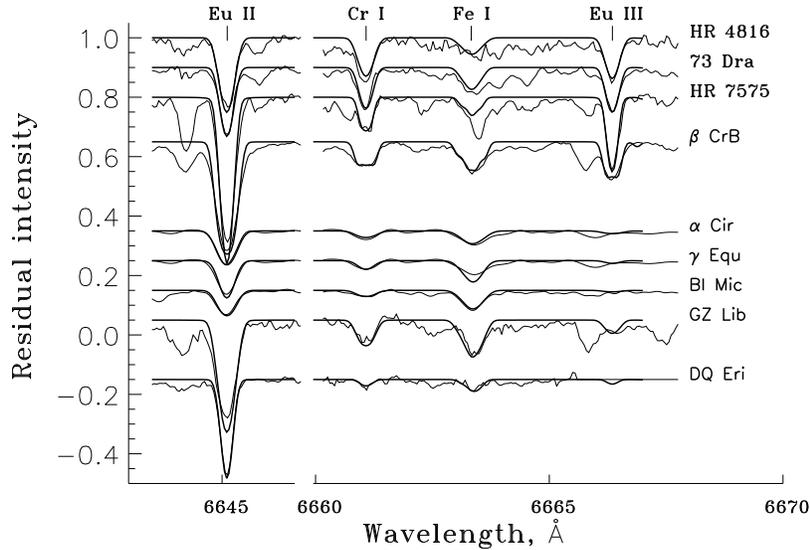,width=11.5cm}
%        \epsfxsize = 115mm
%        \epsffile{eu3_fig.ps}
        \caption{A comparison between the observations and synthesized
      lines of Eu \ii, Eu \iii, Cr \i, and Fe \i.}
\label{Eu_3}
\end{figure}

To synthesize Eu lines properly one needs to know the full hyperfine
splitting for all isotopes. For the Eu\,{\sc ii} $\lambda$ 6645.05 line hfs data
were taken from Biehl (1976). For Eu\,{\sc iii} lines hyperfine splitting is
unknown, therefore we neglect this effect in our study. Synthetic spectrum
calculations which took into account the presence of a magnetic field,
full Zeeman patterns for all synthesized lines and
nine hyperfine components for the Eu\,{\sc ii} line were carried out with the
help of the new code SYNTHMAG. Besides Eu\,{\sc ii} and Eu\,{\sc iii} lines we
also synthesized lines of Cr\,{\sc i} and Fe\,{\sc i} to derive chromium and
iron abundances.
The adopted effective temperatures and magnetic field strengths are
given in Table 1. The synthetic spectra which give the best fit to the
observations are shown in Fig. 1 by thick lines. Observations are
shown by thin lines.

\section{Results}

The results on the Cr, Fe and Eu abundances in the atmospheres of the programme
stars are presented in Table 1.
One immediately sees that
non-roAp CP2 stars are more chemically peculiar then their roAp
counterparts. It is the main conclusion of the present paper, which still
needs confirmation based on the observations of a larger sample of stars
of both groups.

Assuming an ionization balance in the stellar atmospheres we estimated the
astrophysical oscillator strength of the Eu\,{\sc iii} $\lambda$ 6666.347 line.
The mean value obtained from individual estimations given in the last
column of Table 1 results in log(gf)=1.10 $\pm$ 0.10.
It gives us an upper limit for the
oscillator strength because we did not take hfs for this line into
account.
 
\acknowledgements
% Do not leave a blank line here! <---------------------->
International cooperation was supported by the Fonds zur F\"orderung der
Wissenschaftlichen Forschung (project S7303-AMS) and by the Russian Federal
program ``Astronomy'' (grant 2.2.1.5).


\begin{thebibliography}{}
\bibitem{}
  Biehl D. 1976, PhD thesis, Christian-Albrechts-Universit\"at, Institut 
  f\"ur Physik und Sternwarte, pp.~187
\bibitem{}
  Piskunov N. E., Kupka F., Ryabchikova T. A., Weiss W. W.,
  Jeffrey C.S. 1995, A\&AS 112, 525
\bibitem{}
  Sugar J., Spector N. 1974, JOSA 64, 1484
\end{thebibliography}
\end{document}